\documentstyle[aasms4,12pt]{article}

\lefthead{Shimada et al.}
\righthead{The Nuclear Activity of the Galaxies in the Hickson Compact Groups}

\begin{document}

\title{THE NUCLEAR ACTIVITY OF THE GALAXIES IN THE HICKSON COMPACT GROUPS}

\author{Masashi Shimada, Youichi Ohyama, Shingo Nishiura, Takashi Murayama, 
\& Yoshiaki Taniguchi}

\affil{Astronomical Institute, Graduate School of Science, 
       Tohoku University, Aoba, Sendai 980-8578, Japan}

\begin{abstract}
In order to investigate the nuclear activity of galaxies residing 
in compact groups of galaxies,
we present results of our optical spectroscopic program made at 
Okayama Astrophysical Observatory.
We have performed optical spectroscopy of 69 galaxies which belong
to 31 Hickson Compact Groups (HCGs)  of Galaxies.
Among them, three galaxies
have discordant redshifts. Further, spectral quality is too poor to classify
other three galaxies. Therefore, we describe our results for the remaining 
63 galaxies.

Our main results are summarized below.
(1) We have found in our sample; 28 AGN,
16 \ion{H}{2} nuclei, and 19 normal galaxies which show no emission line.
We used this HCG sample for statistical analyses.
(2) Comparing the frequency distributions of activity types
between the HCGs and the field galaxies whose data are taken 
from Ho, Filippenko, \&
Sargent (382 field galaxies), we find that the frequency of \ion{H}{2}
nuclei in the HCGs is significantly less than that in the field.
However, this difference may be due to selection bias that our HCG 
sample contains more early-type galaxies than the field, 
because it is known that \ion{H}{2} nuclei are rarer in early-type 
galaxies than in later ones.
(3) Applying correction this morphological bias to the HCG sample,
we find that there is no statistically significant difference in the frequency of
occurrence of emission-line galaxies between the HCGs and the field.
This implies that the dense galaxy environment in the HCGs does not
affect triggering both the AGN activity and the nuclear starburst.
We discuss some implications on the nuclear activity in the HCG galaxies.
\end{abstract}


\section{INTRODUCTION}

It is known that compact groups of galaxies provide the densest 
galaxy environment rather than binary galaxies, loose groups of 
galaxies and clusters of galaxies (Hickson 1982; Hickson et al.\ 1992). 
Therefore, frequent galaxy collisions are expected to trigger 
either some nuclear activity or intense star formation
in their member galaxies (Hickson et al.\ 1989; Zepf, Whitmore, \& Levison 1991;
Zepf \& Whitmore 1991; Zepf 1993; Verdes-Montenegro et al.\ 1998). 
Further, compact groups would evolve into 
other populations in the universe because they would be able to merge into one
stellar system within a timescale shorter than the Hubble time
(Hickson et al.\ 1992; Barnes 1989; Weil \& Hernquist 1996). 
Indeed previous studies have shown possible evidence that
galaxy collisions may trigger either the nuclear activity or starbursts
in the HCGs; e.g., HCG 16 (Ribeiro et al.\ 1996; de Carbalho \& Coziol 1999),
HCG 31 (Iglesias-P\'aramo \& V\'{\i}lchez 1997a),
HCG 62 (Valluri \& Anupama 1996), HCG 90 (Longo et al.\ 1995), and
HCG 95 (Iglesias-P\'aramo \& V\'{\i}lchez 1997b).

On the other hand, other statistical studies have shown that there may be no strong
evidence for the unusually enhanced activity in the HCGs.
Hickson et al.\ (1989) found that the far-infrared (FIR) emission is 
enhanced in the HCGs. However, later careful analysis
of FIR data of HCGs showed that there is no firm evidence for the enhanced
FIR emission in the HCGs (Sulentic \& de Mello Rabaca 1993). 
Radio continuum properties of the HCG galaxies do not show evidence for
the enhanced nuclear activity  with respect to field spiral galaxies
although the radio continuum emission from the nuclear region 
tends to be stronger than that from field spirals (Menon 1992, 1995).

More recently, Coziol et al.\ (1998) have shown from a spectroscopic
survey for 17 HCGs (de Carvalho et al.\ 1997) that active galactic
nuclei (AGN) are preferentially located in the most early-type and
luminous members in the HCGs. This result suggests possible relations 
among  activity types, morphologies, and densities of galaxies in HCGs.
V\'{\i}lchez \& Iglesias-P\'aramo (1998a) made an H$\alpha$ emission 
imaging survey for a sample of HCGs and found that over 85\% of the 
early-type galaxies in their sample were detected in H$\alpha$
(V\'{\i}lchez \& Iglesias-P\'aramo 1998b). However, they interpreted that
the excess emission in H$\alpha$ is attributed to photoionization by
massive stars rather than AGN.
Therefore, it is still uncertain what kind of activity is 
preferentially induced in the nuclear regions of HCG galaxies. 

In order to investigate nuclear emission-line activity
of HCG galaxies in detail,
our attention is again addressed to an investigation on how 
frequent galaxy collisions are related to the occurrence of both nuclear 
activity and star-formation activity in HCG galaxies.
In this paper, we present results of our optical spectroscopic
program for a sample of 69 galaxies belonging to 31 HCGs which are randomly
selected in the list of HCG (Hickson 1982).
In the original catalog of HCG (Hickson 1982), 100 compact groups with 
493 galaxies are entried. However, eight groups are now dropped out from
the original sample because they do not have more than two galaxies 
whose redshifts are
accordant (Hickson et al.\ 1989; Hickson 1993; see also Sulentic 1997). 
Therefore, our sample is selected from the remaining 92 HCGs.

\section{OBSERVATIONS}

We have performed optical spectroscopy of 69 galaxies 
in the 31 groups (see Table 1).
The spectroscopic observations were made at Okayama Astrophysical Observatory 
(OAO) 188 cm telescope with the new Cassegrain spectrograph and an SITe 
512$\times$512 CCD camera during a period between 1996 February and 1997 January.
The slit dimension was 1.8 arcsec (width) $\times$ 5 arcmin (length). 
Two-pixel binning was made along the slit and thus the spatial resolution was 
1.75 arcsec per element.
The 600 grooves mm$^{-1}$ grating was used to cover 6300 -- 7050 {\AA} region 
with the spectral resolution of 3.4 {\AA} ($\simeq$ 157 km s$^{-1}$ in 
velocity at 6500 \AA). 
The observations were made under photometric conditions.
The typical seeing during the runs was 2 arcsec.

The data were analyzed using IRAF\footnote{%
Image Reduction and Analysis Facility (IRAF) is
distributed by the National Optical Astronomy Observatories,
which are operated by the Association of Universities for Research
in Astronomy, Inc., under cooperative agreement with the National
Science Foundation.}.
We also used a special data reduction package,
SNGRED (Kosugi et al.\ 1995), developed for OAO Cassegrain spectrograph data.
The reduction was made with a standard
procedure; bias subtraction, flat fielding with the data of the dome flats,
and cosmic ray removal.
Flux calibration was obtained using standard stars available in IRAF.
The nuclear spectra were extracted for individual galaxies with
a 1$\farcs$8 $\times$ 1$\farcs$75 aperture.
The extracted nuclear spectra are shown in Figure 1.
A journal of the observations is given in Table 1.
We also give morphological types of galaxies taken from Hickson (1993;
see also Hickson, Kindle, \& Huchra 1988; Mendes de Oliveira \& Hickson 1994)
and de Vaucouleurs et al.\ (1991) in Table 1.

\section{RESULTS}

\subsection{Classification of Emission-line Activity}

In usual classification schemes for emission-line galaxies, 
some combinations of two emission-line intensity ratios 
(e.g., [\ion{O}{3}]$\lambda$5007/H$\beta$
versus [\ion{N}{2}]$\lambda$6583/H$\alpha$) are often used 
(Veilleux \& Osterbrock 1987).
However, since our spectroscopic program was originally devoted to finding 
kinematical peculiarity of HCG galaxies (Nishiura et al.\ 1999), 
our nuclear spectra cover only a wavelength range between 
6300 -- 7050 \AA. Therefore, emission lines available for the classification 
of nuclear activities are [\ion{O}{1}]$\lambda$6300, 
[\ion{N}{2}]$\lambda\lambda$6548,6583,
H$\alpha$, and [\ion{S}{2}]$\lambda\lambda$6717,6731. 
Among several combinations between a couple of the emission lines listed above,
the most reliable indicator to classify nuclear activities seems the
[\ion{N}{2}]$\lambda$6583/H$\alpha$ ratio (hereafter 
[\ion{N}{2}]/H$\alpha$). 
In fact, Ho, Filippenko, \& Sargent (1997)
showed from the spectroscopic analysis of more than 300 nearby galaxies
that this ratio is useful in distinguishing between AGN and \ion{H}{2} nuclei;
i.e., [\ion{N}{2}]/H$\alpha \geq$ 0.6 for AGN while [\ion{N}{2}]/H$\alpha <$ 
0.6 for \ion{H}{2} nuclei. Therefore, applying this criterion,
we classify the emission-line activity of our HCG galaxies.
Galaxies without emission are referred as ``Abs''; i.e., only
stellar absorption features are seen in the optical spectra.
For eight galaxies, we detected only [\ion{N}{2}] line emission and did 
not detect H$\alpha$ line emission (HCG 10a, 30b, 37a, 51b, 62a, 68a, 
88a, and 93c). We classify them as AGN. 
The emission line flux data and the results of the classification are
given in Table 2 and Table 3, respectively. .

As shown in Figure 1, some nuclei show evidence for the H$\alpha$ absorption.
Since the H$\alpha$ absorption leads to an underestimation 
of the H$\alpha$ emission,
it would be better to subtract a template spectrum whose
absorption spectral features are nearly the same as those of the concerned 
spectrum from the target galaxy spectrum (see, for example, Ho et al.\ 1997). 
Since, however, we do not have such a template database,
we used the observed [\ion{N}{2}]/H$\alpha$ ratios in our classification.

In particular, in the case of very weak emission-line galaxies,
the H$\alpha$ emission may not be seen if the H$\alpha$
absorption feature is strong. 
The most serious case may be poststarburst galaxies which show very strong
Balmer absorption (e.g., Taniguchi et al.\ 1996 and references therein).
Poststarburst galaxies have H$\alpha$ absorption equivalent widths, 
$EW$(H$\alpha$) $\geq$ 3 \AA. However, the galaxies with H$\alpha$
absorption in our sample have $EW$(H$\alpha$) $\leq$ 2 \AA; i.e.,
our sample contains no conspicuous poststarburst galaxy.
We therefore expect that our emission-line classification is not affected 
by the effect of H$\alpha$ absorption seriously.

Recently, Coziol et al.\ (1998) studied the 
nuclear activity of southern HCG galaxies.
They obtained optical spectra of 82 brightest galaxies in a sample of 
17 HCGs (de Carvalho et al.\ 1997). 
Among the 82 galaxies, 40 galaxies are original HCG
members identified by Hickson (1982). 
Although their sample is taken from the HCGs located in the southern hemisphere,
13 galaxies in their sample were also observed by us. 
Since they used the template subtraction method in their classification of 
nuclear activity, their classification seems to be more reliable than ours. 
In order to examine how our classification based on
the [\ion{N}{2}]/H$\alpha$ ratio without absorption correction 
is reliable, we compare our results with 
those of Coziol et al.\ (1998). The basic data of the 13 HCG 
galaxies commonly observed by both Coziol et al. (1998) and us 
are summarized in Table 4.
We find that both studies give the same activity types for 
late-type galaxies. However, for early-type galaxies, though we 
have classified three galaxies (HCG 40a, 42a, and 87b) as absorption 
galaxies, they classified them AGNs (dwarf LINERs). 
These differences appear attributed to that we do not apply the template
subtraction method while they did.
However, it is noted that all the three galaxies are not typical Seyfert
nuclei but dwarf LINER nuclei. Although our analysis may not miss
typical Seyfert nuclei, it is safe to mention that about a half
(e.g., 3/7 $\simeq$ 43 \%)  of  early type galaxies classified 
absorption galaxies  in our study are AGNs.
This point will be taken into account in later discussion.

Finally, we classified 63 of 69 galaxies we observed as 28 AGNs, 
16 \ion{H}{2} nuclei, and 19 no line emissions. 
Three of the remaining six galaxies are redshift-discordant galaxies 
(HCG 73a, 87d, and 92a). For the other three, the signal-to-noise ratio 
of their spectrum are too low to classify (HCG 34a, 42b, and 52a).
We exclude these six galaxies from the sample in later statistical 
analyses. 

\subsection{Nuclear Activity versus Group Properties}

Although the selection of HCGs was made homogeneously with the above 
criteria, it is known that
the dynamical properties are different from HCG to HCG
(Hickson et al.\ 1992). Therefore, it is interesting to compare 
the nuclear activity in the member galaxies with the dynamical properties
of the groups. 

As we mentioned previously, we adopt the [\ion{N}{2}]/H$\alpha$ 
intensity ratio as a measure of the nuclear activity. 
Since it is known that the nuclear activity type depends on 
the morphological types of host galaxies (e.g., Ho et al.\ 1997); 
i.e., AGN favors early-type galaxies while star-formation activity 
favors later-type ones, it is necessary to investigate 
relationships between the nuclear activity and the group properties
for each morphological type. However, it is generally difficult to 
classify the morphology of galaxies which are interacting 
with their partner(s) (Mendes de Oliveira \& Hickson 1994). 
Therefore, although we give detailed morphological types
for the member galaxies in our sample in Table 1, we classify them 
broadly into the following three classes; 1) early-type galaxies (E/S0), 
2) early-type spirals (S0a -- Sbc), and 3) late-type spirals 
(Sc or later). In Figures 2 -- 4, we show diagrams of 
[\ion{N}{2}]/H$\alpha$ against the number density of the groups 
$\rho_{\rm N}$ (Hickson et al. 1992), the radial velocity dispersion 
of the groups $\sigma_{\rm r}$ (Hickson et al. 1992), and the crossing 
time of the groups $t_{\rm c}$ (Hickson et al. 1992), respectively. 
We adopt the null hypothesis that the [\ion{N}{2}]/H$\alpha$ ratio 
is correlated with each dynamical parameter and apply 
the Spearman-rank statistical test for all the correlations shown 
in Figures 2, 3, and 4. A summary of the statistical tests is given 
in Table 5. We find that there is no statistically significant 
correlation. Therefore, it is concluded that the nuclear activity of 
galaxies studied here has no physical relation to the dynamical 
properties of the groups.
For disk galaxies in nearby HCGs, Iglesias-P\'{a}ramo \& V\'{i}lchez 
(1999) have found no clear correlations between 
the $L_{\rm H\alpha}/L_{\rm B}$ ratio and the dynamical properties
of the groups. 
Our results are consistent with their results.  

\subsection{Comparison of the Nuclear Activity between the HCG galaxies 
and Field Galaxies}

Our spectroscopic analysis shows that AGN is found in almost half
of the HCG galaxies and star-forming activity is found in a
quarter of the sample. An important question arises  as whether 
or not these frequencies are unusual with respect to those 
in environment with less galaxy collisions. 
In order to examine this issue, we at first make a control sample
which consists of so-called field galaxies and then compare the 
nuclear activity between the HCG galaxies and the field galaxies. 

Recently Ho et al. (1995, 1997) have made an
extensive spectroscopic survey for nearby galaxies using the Palomar
Observatory 5 m telescope. Their sample contains 486 galaxies with 
$B_{T} \leq 12.5$ and $\delta > 0^\circ$ where $B_{T}$ is 
the apparent total $B$
magnitude and $\delta$ is the declination. 
In order to make a sample of field galaxies, we have omitted the following 
galaxies from their sample; 1) galaxies belong to the Virgo cluster,
2) binary/interacting galaxies, 3) HCG galaxies (HCG 44a = NGC 3190,
HCG 44b = NGC 3193, HCG 44c = NGC 3185, HCG 61a = NGC 4169, 
HCG 68a = NGC 5353, and HCG 68b = NGC 5354), 4) NGC 1003 whose 
activity type is uncertain, and 5) the Hubble type is uncertain for
five galaxies (NGC 63, 812, 2342, 7798, and UGC 3714).
Excluding the above galaxies, we obtain a sample of 382 field galaxies
which consist of 167 AGNs, 174 \ion{H}{2} nuclei, 
and 41 normal galaxies. 
This sample has no matching to the HCG sample in both 
apparent magnitude and morphology.
Since the majority of the HCG galaxies are fainter than the field
galaxies observed by Ho et al.\ (1997), it is difficult to obtain a
magnitude-matched sample of field galaxies.
However, when we compare the nuclear activity between the HCG
galaxies and the field galaxies, we will take account of the morphological
difference between the HCGs and the fields.

In Figure 5, we show the frequency distributions of activity types
for the HCGs (upper panels) and for the field (lower panels).
Applying the $\chi^2$ test, we examine whether or not the frequency
distributions of the activity types for the HCGs are significantly
different from those for the field galaxies for the morphological 
samples of E -- S0, S0a -- Sbc, Sc or later, and all the galaxies 
(the total sample).
We adopt the null hypothesis that the HCG galaxies and field galaxies 
come from the same underlying distribution of the activity types.
The results of our statistical test are summarized in Table 6.
Although the difference in the frequency distribution is not 
statistically significant for each morphological type, 
the difference for the total sample is significant in that the HCGs
have less \ion{H}{2} nuclei while have more absorption galaxies
than the field galaxies. The \ion{H}{2} nuclei and the absorption 
galaxies are found in 26\% and 31\% of the HCG galaxies, respectively. 
On the other hand, in the field, the \ion{H}{2} nuclei 
share 46\% while the absorption galaxies share only 11\% of the sample.

Taking account that the nuclear activity type depends on
the morphological types of host galaxies (e.g., Ho et al.\ 1997),
we examine the difference in the morphological type distribution
between the HCG galaxies and the field ones. 
In Figure 6, we show the frequency distributions
of morphological types for each activity type and for the total sample.
Applying the $\chi^2$ test, we examine whether or not the frequency
distributions of the morphological types for our HCG galaxies are 
significantly different from those for the field galaxies for the 
nuclear activity type of
AGN, \ion{H}{2}, absorption, and the total sample.
We adopt the null hypothesis that
the HCG galaxies and field galaxies
come from the same underlying distribution of the morphological types.
The results of our statistical test are summarized in Table 7. 
Our HCG sample contain more E -- S0 galaxies while less late-type 
spirals than the field. This leads to the under population of 
\ion{H}{2} nuclei in the HCG sample because \ion{H}{2} nuclei favor 
such late-type spirals. 
However, the frequency of occurrence of AGN in the HCGs is nearly
the same as that in the field. 
A remarkable difference may be that the \ion{H}{2} nuclei are 
found in E -- S0 galaxies more frequently in the HCGs ($\simeq$ 13\%) 
than in the field ($\simeq$ 2\%).
This result appears consistent with the finding by Zepf et al.\ (1991);
there are a number of early-type galaxies with unusually blue colors,
suggesting the enhanced star formation in early type galaxies.

We have found some interesting difference in the frequency distributions
of the activity types between the HCGs and the field described above. 
However, since the frequency distribution of morphological types 
is different between the two samples, we cannot conclude that the
differences are real. In order to check the effect of the difference 
in the morphological type distributions, we estimate the frequency of
AGN, \ion{H}{2} nuclei, and absorption galaxies in the HCGs if 
the morphological type distribution in the HCGs is the same  as that 
in the field. For example, we can estimate the expected number of AGN 
in the HCGs as 
$N_{\rm AGN}^{\rm exp}({\rm HCG}) 
= N_{\rm E-S0} \times P_{\rm AGN, E-S0}({\rm Field}) +
N_{\rm S0a-Sbc} \times P_{\rm AGN, S0a-Sbc}({\rm Field}) +
N_{\rm Sc} \times P_{\rm AGN, Sc}({\rm Field})$
where $P_{x, y}({\rm Field})$ is the probability that galaxies with
the morphological type $y$ have the activity type $x$ in the field sample.
We can also estimate both $N_{\rm HII}^{\rm exp}({\rm HCG})$ and 
$N_{\rm Abs}^{\rm exp}({\rm HCG})$ in a similar way.
We also adopt the null hypothesis that the observed distribution
is  the same as the expected distribution and
apply the $\chi^2$ test.
The results are given in Table 8.
We find that there is no statistical difference
in the activity-type distributions between the HCGs and the field.
Hence, we conclude that the nuclear activity in the HCGs is not different
from that in the field under the assumption that the morphology-activity
relation is the same between the HCGs and the field.

As mentioned in section 3.1, our spectral analysis may miss 
dwarf LINERs roughly in a half of early-type galaxies studied here.
If we assume that a half of the early-type galaxies classified as
^^ ^^ Abs" could be AGNs,
our 63 HCG galaxies are classified 
into 36 AGNs, 16 \ion{H}{2} nuclei, and 19 Abs.
In this case, we obtain
$P(\chi^{2})=0.50$. This means that the activity distribution of 
HCG galaxies is again indistinguishable from that of the field galaxies. 

\section{DISCUSSION}

Our main results are summarized below.
(1) We have described the results of our spectroscopic program for a 
sample of 63 galaxies in the 28 HCGs.
We have found in our sample; 28 AGN,
16 \ion{H}{2} nuclei, and 19 normal galaxies which show no emission line.
We used this HCG sample for statistical analyses.
(2) Comparing the frequency distributions of activity types
between the HCGs and the field whose data are taken from Ho, Filippenko,
\& Sargent (382 field galaxies),
we find that the frequency of occurrence of \ion{H}{2} nuclei in the HCGs
is significantly less than that in the field.
However, our HCG sample contains more early-type galaxies than the field,
the above difference for the \ion{H}{2} nuclei may be due to this morphology
bias because it is known that \ion{H}{2} nuclei are rarer in early-type
galaxies than in later ones.
(3) Correcting this morphological bias to the HCG sample,
we find that there is no significant difference in the frequency of
occurrence of emission-line galaxies between the HCGs and the field.
This implies that the dense galaxy environment in the HCGs does not
affect triggering both AGNs and nuclear starbursts.
(4) Since our classification of nuclear activities are judged
by the raw optical spectra, we may miss some less-luminous AGNs,
in particular in early-type galaxies. Even though this effect is
taken into account, the distributions of activity types of HCG
galaxies are indistinguishable from those of field galaxies.

Our finding seems surprising because it is widely accepted that 
galaxy interactions lead to either nuclear activity such as AGN or
nuclear starbursts or both (see for a review Shlosman, Begelman, 
\& Frank 1990; Barnes \& Hernquist 1992).
Indeed, in 1980's, several systematic observational investigations of
interacting or binary galaxies suggested that galaxy collisions may raise
both nuclear activity and intense star formation
(e.g., Kennicutt et al.\ 1984; Keel et al.\ 1985; Dahari 1985; 
Bushouse 1986, 1987) although the statistical significance was not so high;
i.e., $\simeq$ 90 -- 95\% (see for recent papers; 
De Robertis, Yee, \& Hayhoe 1998; Taniguchi 1999).
In addition, luminous and ultraluminous infrared galaxies are often detected
in strongly interacting galaxies and merging galaxies (Sanders et al.\ 1988;
see for a review Sanders \& Mirabel 1996).
Numerical simulations of interacting or merging galaxies have shown that
gas fueling driven by galaxy interaction occurs efficiently
(e.g., Noguchi 1988; Olson \& Kwan 1990a, 1990b; Mihos \& Hernquist 1994b).

If tidal interactions lead to the formation of AGN and/or nuclear starbursts,
we would observe a large number of such active galaxies in the HCGs 
because the member galaxies are expected to have experienced many 
tidal interactions during the course of their dynamical evolution.
Galaxy interactions affect the star formation activity in galactic
disks because the effect of tidal interactions is much stronger in the
outer parts than in the nuclear regions (e.g., Noguchi \& Ishibashi 1986;
see also Kennicutt et al.\ 1987). Indeed, some radio studies have 
revealed that a large fraction of HCG spirals are \ion{H}{1} deficient 
(Williams \& Rood 1987; Huchtmeier 1997). 
If a HCG contains several gas-rich spiral galaxies, the average star 
formation rate would be more enhanced than that in the field galaxies
(e.g., Young et al.\ 1986). However, such excess has not yet been 
confirmed by IRAS observations (Sulentic \& De Mello Rabaca 1993 
and references therein).
Although deficient of the atomic hydrogen gas in HCG spirals implies 
that it is expected to occur intense star formation in HCG galaxies, 
Moles et al. (1994) have concluded that there are no strong 
starbursting galaxies in HCGs by optical and infrared observations. 
These results indicate that frequent galaxy collisions are not always 
able to increase the star formation rate intensely. 
Although no enhancement of far infrared emission may be partly 
attributed to that the HCGs prefer 
early-type spiral galaxies as well as elliptical ones,
it should be noted that roughly half galaxies in the HCGs are 
late-type spirals and irregular galaxies (Hickson et al.\ 1988;
Mendes de Oliveira \& Hickson 1994). 
Therefore, it is suggested that off-nuclear star-formation activity is 
also not enhanced in the HCGs with respect to field galaxies. 

Coziol et al.\ (1998) have shown from a spectroscopic
survey for 17 HCGs (de Carvalho et al.\ 1997) that
AGN are preferentially located in the most early-type and
luminous members in the HCGs, suggesting a correlation between
activity types, morphologies, and densities of galaxies in HCGs.
They searched more possible member galaxies outside 
the original HCG members and then found the above interesting
observational properties. However, our spectroscopic survey
was made for only the original HCG members. Therefore, we do not
think that our results are inconsistent with their results.
An interesting point suggested by Coziol et al.\ (1998) is that
AGN is preferentially found in luminous, early-type galaxies.
Verdes-Montenegro et al.\ (1998) showed from their $^{12}$CO($J$=1-0)
emission survey for a large number of HCG galaxies that a number of 
early-type galaxies are detected in CO as well as in FIR. 
In addition, early-type galaxies with unusually blue colors are 
also found by Zepf et al.\ (1991). Although all these may be still
circumstantial lines of evidence, it is suggested that the majority
of early-type galaxies in the HCGs are affected by some environmental
effect. One possible important effect is a merger between an early-type
galaxy and a gas-rich galaxy such as a late-type spiral or a small
satellite galaxy since galaxy mergers between unequal galaxies
may lead to the formation of S0 galaxies (Bekki 1998).

The above arguments suggest that mere tidal interactions between
galaxies are not responsible for the triggering intense nuclear
activities. Recently, instead of  mere tidal interactions,
minor mergers have been appreciated as a more important  
triggering mechanism both for nuclear starbursts
(Mihos \& Hernquist 1994a; Hernquist \& Mihos 1995;
Taniguchi \& Wada 1996) and for Seyfert nuclei (De Robertis
et al. 1998; Taniguchi 1999; see for an earlier indication
Gaskell 1985). If this is the case, it is not surprising that
the nuclear activities in the HCG galaxies are not significantly
different from those in the field galaxies.
Furthermore, if major mergers are more important to activate 
mode luminous starbursts and AGNs (e.g., Sanders et al. 1988),
it is suggested that most of the HCGs have not yet experienced
such major mergers in the member galaxies. 
Since the dynamical relaxation timescale for the HCGs is 
shorter than the Hubble times, it is expected that each HCG
will merge into one within a timescale of several Gyr
(Hickson et al. 1992). Therefore, the HCG are expected to evolve 
either to
luminous or ultraluminous infrared galaxies via multiple mergers
(Xia et al. 1997; Taniguchi, Wada, \& Murayama 1997;
Taniguchi \& Shioya 1998; L\'ipari et al. 2000;
Borne et al. 2000), or to quasars (Sanders et al. 1988; Taniguchi, 
Ikeuchi, \& Shioya 1999) or to ordinary-looking elliptical 
galaxies (Barnes 1989; Weil \& Hernquist 1996; Nishiura et al. 1997).

\begin{acknowledgements}
We are grateful to the staff of OAO for kind help of the observations.
We would like to thank an anonymous referee for useful comments and
suggestions.
YO and TM are JSPS Fellows.
This work was partly supported by the Ministry of Education, Science, Culture, 
and Sports (Nos. 07044054, 10044052, and 10304013).
\end{acknowledgements}

%----------------------------------------------------------------------
%           References
%----------------------------------------------------------------------

%---------------------------------------------------------------------------------

\newpage

\figcaption{Nuclear spectra of the HCG galaxies.}
\label{fig1}

\figcaption{Correlations between the $[$N {\sc ii}$]$/H$\alpha$ ratio and 
the number density of galaxies in the HCGs $\rho_{\rm N}$. The 
results are shown for all the sample
(top left), E -- S0 galaxies (top right), S0a -- Sbc galaxies (bottom
left), and Sc or later (bottom right).}
\label{fig2}

\figcaption{Correlations between the $[$N {\sc ii}$]$/H$\alpha$ ratio and 
the radial velocity dispersion of galaxies in the HCGs 
$\sigma_{\rm r}$. The results are shown for all the sample
(top left), E -- S0 galaxies (top right), S0a -- Sbc galaxies (bottom
left), and Sc or later (bottom right).}
\label{fig3}

\figcaption{Correlations between the $[$N {\sc ii}$]$/H$\alpha$ ratio and 
the crossing time of galaxies in the HCGs $t_{\rm c}$. The results are 
shown for all the sample
(top left), E -- S0 galaxies (top right), S0a -- Sbc galaxies (bottom
left), and Sc or later (bottom right).}
\label{fig4}

\figcaption{Comparison of frequency distributions of nuclear activity types 
between the HCGs and the field
for all the sample, E -- S0 galaxies, S0a -- Sbc galaxies,
and Sc or later.}
\label{fig5}

\figcaption{Comparison of frequency distributions of morphological types
between the HCGs and the field
for all the sample, AGN, H {\sc ii} nuclei,
and absorption galaxies.}
\label{fig6}

\end{document}

%######################### TABLE 1 ############################################
\begin{deluxetable}{ccrrccc}
\tablenum{1}
\tablecaption{\ A journal of observations}
\tablehead{
\colhead{HCG}     & \colhead{Date} & \colhead{Exp.}
   & \colhead{PA} & \colhead{Type}
   & \colhead{Type} & \colhead{Type} \nl
\colhead{} & \colhead{}  & \colhead{(sec)}
   & \colhead{(deg)} & \colhead{Hickson}
   & \colhead{RC3}   & \colhead{Adopted}
}
\startdata
7a   & 19 Aug 1996&   1800 &  159 & Sb      & (R')SB(r)a:      & Sa  \nl
10a  & 21 Feb 1996&   1800 &   62 & SBb     & SB(r)b           & Sb  \nl
10b  & 07 Jan 1997&   1800 &  153 & E1      & S0$^{-}$:        & S0  \nl
30a  & 21 Feb 1996&   1800 &  124 & SBa     & (R')SB(rs)0$^{+}$: & S0 \nl
30b  & 23 Feb 1996&   1800 &   29 & Sa      & (R')SAB(rs)0$^{+}$: & S \nl
31a  & 09 Jan 1997&   1800 &   80 & Sdm     & Pec              & P    \nl
31b  & 09 Jan 1997&   1800 &   41 & Sm      & S?               & Sm   \nl
31c  & 09 Jan 1997&    860 &   41 & Im      & S?               & Im   \nl
34a  & 23 Feb 1996&   1800 &   84 & E2      & S0?              & S0   \nl
37a  & 22 Feb 1996&   1800 &  171 & E7      & Sb               & Sb   \nl
37b  & 21 Feb 1996&   1800 &   77 & Sbc     & S?               & Sbc  \nl
38a  & 06 Jan 1997&   1800 &  340 & Sbc     & S?               & Sbc  \nl
38b  & 23 Feb 1996&   1800 &   85 & SBd     & S?               & Sd   \nl
38c  & 23 Feb 1996&   1800 &   85 & Im      & S?               & Im   \nl
40a  & 23 Feb 1996&   1800 &  185 & E3      & E                & E    \nl
40b  & 23 Feb 1996&   1800 &   42 & S0      & SA(r)0$^{-}$: pec & S0  \nl
40c  & 23 Feb 1996&   1800 &  122 & Sbc     & SB(rs)b pec    & Sb   \nl
42a  & 09 Jan 1997&   1800 &  134 & E3      & E3:              & E    \nl
42b  & 09 Jan 1997&    240 &   66 & SB0     & SB(rs)0$^{0}$    & S0   \nl
42c  & 09 Jan 1997&   1800 &  134 & E2      & E?               & E    \nl
44a  & 21 Feb 1996&   1800 &  125 & Sa      & SA(s)a pec    & Sa   \nl
44b  & 22 Feb 1996&   1800 &  185 & E2      & E2               & E    \nl
44c  & 21 Feb 1996&   1800 &  130 & SBc     & (R)SB(r)a        & Sa   \nl
44d  & 21 Feb 1996&   1800 &  125 & Sd      & SB(s)c pec       & Sc   \nl
47a  & 21 Feb 1996&   1800 &   15 & SBb     & SA(r):           & Sb   \nl
51a  & 23 Feb 1996&   1800 &  358 & E1      & E                & E    \nl
51b  & 23 Feb 1996&   1800 &  183 & SBbc    & S?               & Sbc  \nl
51f  & 23 Feb 1996&   1800 &  358 & S0      & S0?              & S0   \nl
52a  & 20 Feb 1996&    750 &   90 & SBab    & SB?              & Sab  \nl
53a  & 20 Feb 1996&   1800 &   93 & SBbc    & SB?              & Sbc  \nl
57a  & 25 Feb 1996&   1800 &  108 & Sb      & Sab  pec       & Sab  \nl
61a  & 22 Feb 1996&   1800 &  159 & S0a     & S0               & S0   \nl
61c  & 21 Feb 1996&   1800 &  130 & Sbc     & S                & Sbc  \nl
61d  & 21 Feb 1996&   1800 &   50 & S0      & S?               & S0   \nl
62a  & 22 Feb 1996&   1800 &  130 & E3      & S0$^{+}$       & S0   \nl
62b  & 22 Feb 1996&   1800 &  130 & S0      & S0$^{0}$ pec     & S0   \nl
62c  & 22 Feb 1996&   1800 &  173 & S0      & E+               & E    \nl
67b  & 21 Feb 1996&    900 &   21 & Sc      & Sb             & Sb   \nl
68a  & 22 Feb 1996&   1800 &    0 & S0      & S0             & S0   \nl
68b  & 22 Feb 1996&   1800 &    0 & E2      & S0             & S0   \nl
68c  & 21 Feb 1996&   1800 &   18 & SBbc    & SB(r)b           & Sb   \nl
71a  & 20 Feb 1996&   1800 &   23 & SBc     & Scd:             & Scd  \nl
72a  & 25 Feb 1996&   1800 &  102 & Sa      & S0?              & S0   \nl
73a  & 25 Feb 1996&   1800 &   18 & Scd     & SA(s)c           & Sc   \nl
79a  & 22 Feb 1996&    758 &   63 & E0      & Sa pec           & Sa   \nl
79b  & 15 Aug 1996&   1800 &   70 & S0      & S0?              & S0   \nl
79c  & 18 Aug 1996&   1800 &   34 & S0      & S0 pec           & S0   \nl
79d  & 18 Aug 1996&   1800 &    0 & Sdm     & SB(s)c         & Sc   \nl
80a  & 25 Feb 1996&   1800 &   64 & Sd      & S?               & Sd   \nl
82a  & 15 Aug 1996&   1800 &   37 & E3      & S0               & S0   \nl
82b  & 19 Aug 1996&   1800 &  168 & SBa     & SB0?             & S0   \nl
87a  & 18 Aug 1996&   1800 &   56 & Sbc     & S0$^{0}$ pec   & S0   \nl
87b  & 20 Aug 1996&   1800 &   40 & S0      & S0?              & S0   \nl
87c  & 19 Aug 1996&   1800 &   90 & Sd      & S?               & Sd   \nl
87d  & 20 Aug 1996&   1800 &   40 & Sd      &                  & Sd   \nl
88a  & 15 Aug 1996&   1800 &  132 & Sb      & Sb               & Sb   \nl
88b  & 15 Aug 1996&   1800 &   31 & SBb     & SB(r)a pec:      & Sa   \nl
88c  & 15 Aug 1996&   1800 &   31 & Sc      & SAB(r)bc?        & Sbc  \nl
88d  & 18 Aug 1996&   1800 &   71 & Sc      & S?               & Sc   \nl
89a  & 19 Aug 1996&   1800 &   57 & Sc      & S?               & Sc   \nl
92a  & 15 Aug 1996&   1800 &  137 & Sd      & SA(s)d           & Sd   \nl
92b  & 15 Aug 1996&   1800 &   22 & Sbc     & SB(s)bc pec      & Sbc  \nl
92c  & 18 Aug 1996&   1800 &  140 & SBa     & SB(s)bc pec      & Sbc  \nl
92d  & 15 Aug 1996&   1800 &  109 & SB0     & E2 pec           & E    \nl
92e  & 18 Aug 1996&   1800 &  110 & Sa      &                  & Sa   \nl
93a  & 19 Aug 1996&   1800 &   98 & E1      & SA0$^{-}$        & S0   \nl
93b  & 19 Aug 1996&   1800 &  143 & SBd     & SB(s)cd pec      & Scd  \nl
93c  & 19 Aug 1996&   1800 &   98 & SBa     & (R)SAB(s)0/a pec & S0/a \nl
96a  & 19 Aug 1996&    480 &   90 & Sc      & SA(r)bc pec      & Sbc  \nl
\enddata
\end{deluxetable}

%----------------------------------------------------------------------------
%%%%%%%%%%%%%%%%%%%%% TABLE 2 %%%%%%%%%%%%%%%%%%%%%%%%%%%%%
% Another File, Shimada_table2.tex

%----------------------------------------------------------------------------
\clearpage
%%%%%%%%%%%%%%%%%%%%% TABLE 3 %%%%%%%%%%%%%%%%%%%%%%%%%%%%%
\begin{deluxetable}{cccccccc}
\tablenum{3}
\tablecaption{\ Line Ratio and Dynamical Properties}
\tablehead{
  %%% HEAD 1 %%%
    \colhead{HCG} &
    \colhead{$[$\ion{N}{2}$]$/H$\alpha$\tablenotemark{a}} &
    \colhead{$[$\ion{S}{2}$]$/H$\alpha$\tablenotemark{b}} &
    \colhead{$[$\ion{O}{1}$]$/H$\alpha$\tablenotemark{c}} &
    \colhead{$\log\rho_{\rm N}$} &
    \colhead{$\log\sigma_{\rm r}$} &
    \colhead{$\log H_{0}t_{c}$} &
    \colhead{Activity} \nl
  %%% HEAD 2 %%%
    \colhead{} &
    \colhead{} &
    \colhead{} &
    \colhead{} &
    \colhead{(Mpc$^{-3}$)} &
    \colhead{(km s$^{-1}$)} &
    \colhead{} &
    \colhead{Type}
}

\startdata
 07a &  0.49 $\pm$ 0.07  &  0.24 $\pm$ 0.09  &                 
     &  4.00             &  1.95             & -1.40 & HII  \nl
 10a &                   &                   &
     &  3.07             &  2.32             & -1.48 & AGN  \nl
 10b &                   &                   &
     &  3.07             &  2.32             & -1.48 & Abs  \nl
 30a &                   &                   &
     &  3.85             &  1.86             & -1.22 & Abs  \nl
 30b &                   &                   & 
     &  3.85             &  1.86             & -1.22 & AGN  \nl
 31a &  0.17 $\pm$ 0.02  &  0.28 $\pm$ 0.03  &
     &  6.13             &  1.75             & -1.80 & HII  \nl
 31b &                   &  0.35 $\pm$ 0.08  &
     &  6.13             &  1.75             & -1.80 & HII  \nl
 31c &  0.08 $\pm$ 0.01  &  0.13 $\pm$ 0.01  &  0.02 $\pm$ 0.01
     &  6.13             &  1.75             & -1.80 & HII  \nl
 34a &  1.89 $\pm$ 2.39  &                   &
     &  5.41             &  2.50             & -2.44 & AGN? \nl
 37a &                   &                   &
     &  4.70             &  2.60             & -2.27 & AGN  \nl
 37b &  1.26 $\pm$ 1.10  &                   &
     &  4.70             &  2.60             & -2.27 & AGN  \nl
 38a &  0.34 $\pm$ 0.07  &  0.33 $\pm$ 0.15  & 
     &  3.55             &  1.11             &  0.88 & HII  \nl
 38b &  0.56 $\pm$ 0.06  &  0.14 $\pm$ 0.05  &  
     &  3.55             &  1.11             &  0.88 & HII  \nl
 38c &  0.53 $\pm$ 0.17  &  0.27 $\pm$ 0.14  &
     &  3.55             &  1.11             &  0.88 & HII  \nl
 40a &                   &                   &
     &  5.54             &  2.17             & -2.12 & Abs  \nl
 40b &                   &                   &
     &  5.54             &  2.17             & -2.12 & Abs  \nl
 40c &  0.59 $\pm$ 0.19  &                   & 
     &  5.54             &  2.17             & -2.12 & HII  \nl
 42a &                   &                   &
     &  4.03             &  2.33             & -1.81 & Abs  \nl
 42b &                   &                   &
     &  4.03             &  2.33             & -1.81 &  ?   \nl
 42c &                   &                   &
     &  4.03             &  2.33             & -1.81 & Abs  \nl
 44a & 21.46 $\pm$ 18.45 & 11.11 $\pm$ 10.40 &
     &  4.24             &  2.13             & -1.65 & AGN  \nl
 44b &                   &                   &
     &  4.24             &  2.13             & -1.65 & Abs  \nl
 44c &  0.75 $\pm$ 0.06  &  0.42 $\pm$ 0.05  &
     &  4.24             &  2.13             & -1.65 & AGN  \nl
 44d &  0.32 $\pm$ 0.24  &                   &
     &  4.24             &  2.13             & -1.65 & HII  \nl
 47a &                   &                   &
     &  4.30             &  1.63             &  0.66 & Abs  \nl
 51a &                   &                   &
     &  3.77             &  2.38             & -1.73 & Abs  \nl
 51b &                   &                   &
     &  3.77             &  2.38             & -1.73 & AGN  \nl
 51f &                   &                   &
     &  3.77             &  2.38             & -1.73 & Abs  \nl
 52a &                   &                   &
     &  3.04             &  2.26             & -1.43 &  ?   \nl
 53a &  1.53 $\pm$ 1.15  &                   &
     &  3.58             &  1.91             & -1.14 & AGN  \nl
 57a &  1.39 $\pm$ 0.74  &                   &
     &  3.64             &  2.43             & -1.69 & AGN  \nl
 61a & 10.33 $\pm$ 9.51  &  8.17 $\pm$ 7.79  & 
     &  4.48             &  1.94             & -1.60 & AGN  \nl
 61c &  0.73 $\pm$ 0.11  &  0.33 $\pm$ 0.10  &
     &  4.48             &  1.94             & -1.60 & AGN  \nl
 61d &                   &                   &
     &  4.48             &  1.94             & -1.60 & Abs  \nl
 62a &                   &                   &
     &  4.69             &  2.46             & -2.16 & AGN  \nl
 62b &                   &                   &
     &  4.69             &  2.46             & -2.16 & Abs  \nl
 62c &                   &                   &
     &  4.69             &  2.46             & -2.16 & Abs  \nl
 67b &  3.48 $\pm$ 2.98  &                   &
     &  3.91             &  2.32             & -1.76 & AGN  \nl
 68a &                   &                   &
     &  4.52             &  2.19             & -1.79 & AGN  \nl
 68b &  4.51 $\pm$ 2.98  &                   &
     &  4.52             &  2.19             & -1.79 & AGN  \nl
 68c &  0.77 $\pm$ 0.06  &  0.39 $\pm$ 0.05  &
     &  4.52             &  2.19             & -1.79 & AGN  \nl
 71a &  2.02 $\pm$ 0.25  &                   &
     &  3.76             &  2.62             & -2.04 & AGN  \nl
 72a &  0.92 $\pm$ 0.82  &                   &
     &  4.33             &  2.42             & -2.00 & AGN  \nl
 73a &  0.42 $\pm$ 0.16  &                   &
     &                   &                   &       & HII  \nl
 79a &  0.61 $\pm$ 0.46  &                   &
     &  6.49             &  2.14             & -2.43 & AGN  \nl
 79b &  0.45 $\pm$ 0.07  &  0.27 $\pm$ 0.06  &
     &  6.49             &  2.14             & -2.43 & HII  \nl
 79c &                   &                   &
     &  6.49             &  2.14             & -2.43 & Abs  \nl
 79d &  0.15 $\pm$ 0.07  &  0.50 $\pm$ 0.16  &
     &  6.49             &  2.14             & -2.43 & HII  \nl
 80a &  0.73 $\pm$ 0.11  &                   &
     &  4.78             &  2.43             & -2.16 & AGN  \nl
 82a &                   &                   &
     &  3.43             &  2.79             & -2.07 & Abs  \nl
 82b &  0.57 $\pm$ 0.20  &                   &
     &  3.43             &  2.79             & -2.07 & HII  \nl
 87a &  0.97 $\pm$ 0.57  &                   &
     &  4.39             &  2.08             & -1.56 & AGN  \nl
 87b &                   &                   &
     &  4.39             &  2.08             & -1.56 & Abs  \nl
 87c &  0.43 $\pm$ 0.11  &                   &
     &  4.39             &  2.08             & -1.56 & HII  \nl
 87d &                   &                   &
     &                   &                   &       &  ?   \nl
 88a &                   &                   &
     &  3.49             &  1.43             &  0.94 & AGN  \nl
 88b &  6.47 $\pm$ 3.83  &                   &
     &  3.49             &  1.43             &  0.94 & AGN  \nl
 88c &  0.42 $\pm$ 0.13  &                   &
     &  3.49             &  1.43             &  0.94 & HII  \nl
 88d &  0.47 $\pm$ 0.11  &                   &
     &  3.49             &  1.43             &  0.94 & HII  \nl
 89a &  0.48 $\pm$ 0.13  &                   &
     &  3.67             &  1.74             & -0.84 & HII  \nl
 92a &  0.42 $\pm$ 0.20  &                   &
     &                   &                   &       & HII  \nl
 92b &                   &                   &
     &  4.63             &  2.59             & -2.27 & Abs  \nl
 92c &  1.46 $\pm$ 0.09  &  0.97 $\pm$ 0.10  &  0.28 $\pm$0.05
     &  4.63             &  2.59             & -2.27 & AGN  \nl
 92d &                   &                   & 
     &  4.63             &  2.59             & -2.27 & Abs  \nl
 92e &                   &                   &
     &  4.63             &  2.59             & -2.27 & Abs  \nl
 93a &  3.74 $\pm$ 1.31  &  2.55 $\pm$ 1.03  &
     &  3.43             &  2.32             & -1.59 & AGN  \nl
 93b &  0.66 $\pm$ 0.09  &  0.30 $\pm$ 0.11  &
     &  3.43             &  2.32             & -1.59 & AGN  \nl
 93c &                   &                   &
     &  3.43             &  2.32             & -1.59 & AGN  \nl
 96a &  0.99 $\pm$ 0.21  &  0.54 $\pm$ 0.16  &
     &  4.54             &  2.12             & -1.76 & AGN  \nl
\enddata
\tablenotetext{a}{[\ion{N}{2}]$\lambda$6583/H$\alpha$. 
                  The error is one $\sigma$ value.}
\tablenotetext{b}{[\ion{S}{2}]($\lambda$6717+$\lambda$6731)/H$\alpha$.
                  The error is one $\sigma$ value.} 
\tablenotetext{c}{[\ion{O}{1}]$\lambda$6300/H$\alpha$.
                  The error is one $\sigma$ value.}
\end{deluxetable}

%----------------------------------------------------------------------------
\clearpage
%%%%%%%%%%%%%%%%%%%%%%%%%%%%% TABLE 4 %%%%%%%%%%%%%%%%%%%%%%%%%%%%%
\begin{deluxetable}{cccccc}
\tablenum{4}
\tablewidth{40em}
\tablecaption{\ Comaparison of nuclear activities between our study 
               and Coziol et al. (1998)
             } 
\tablehead{
     %%% HEAD 1 %%%
     \colhead{HCG} &
     \colhead{Type} &
     \multicolumn{2}{c}{Coziol et al. (1998)} &
     \multicolumn{2}{c}{Our study} \nl
     %%% HEAD 2 %%%
     \colhead{} &
     \colhead{Adopted}&
     \colhead{$[$\ion{N}{2}$]$/H$\alpha$} &
     \colhead{Activity Type} &
     \colhead{$[$\ion{N}{2}$]$/H$\alpha$} &
     \colhead{Activity Type} 
}
\startdata
40a & E  & 0.87$\pm$0.06 & AGN &               & Abs \nl
40b & S0 &               & Abs &               & Abs \nl 
42a & E  & 1.78$\pm$0.04 & AGN &               & Abs \nl 
42c & E  &               & Abs &               & Abs \nl 
62a & S0 & 0.60$\pm$0.01 & AGN &               & AGN\tablenotemark{a} \nl 
62b & S0 &               & Abs &               & Abs \nl 
62c & E  &               & Abs &               & Abs \nl 
87a & S0 & 0.83          & AGN & 0.97$\pm$0.57 & AGN \nl 
87b & S0 & 1.29          & AGN &               & Abs \nl 
87c & Sd & 0.44          & HII & 0.43$\pm$0.11 & HII \nl 
88a & Sb & 0.89          & AGN &               & AGN\tablenotemark{a} \nl 
88b & Sa & 1.91          & AGN & 6.47$\pm$3.83 & AGN \nl 
88d & Sc & 0.32          & HII & 0.47$\pm$0.11 & HII \nl 

\tablenotetext{a}{We have not detected H$\alpha$ but have detected 
$[$N {\sc II}$]\lambda$6583.}

\enddata
\end{deluxetable}

%----------------------------------------------------------------------------
\clearpage
%%%%%%%%%%%%%%%%%%%%%%%%%%%%% TABLE 5 %%%%%%%%%%%%%%%%%%%%%%%%%%%%%
\begin{deluxetable}{cccc}
\tablenum{5}
\tablewidth{40em}
\tablecaption{\ A summary of the Spearman rank test for 
              the correlations between $[$N {\sc ii}$]$/H$\alpha$ ratio 
              and the dynamical properties of the HCGs\tablenotemark{a}
             } 
\tablehead{
     %%% HEAD 1 %%%
     \colhead{} &
     \multicolumn{3}{c}{log([\ion{N}{2}]$\lambda$6583/H$\alpha$)} \nl
     %%% HEAD 2 %%%
     \colhead{} &
     \colhead{$\log \rho_{\rm N}$}&
     \colhead{$\log \sigma_{\rm r}$} &
     \colhead{$\log H_{\rm 0}t_{\rm c}$}
}
\startdata
All       &   0.82   &   0.004 &   0.97   \nl
\hline
E$-$S0    &   0.43   &   0.93  &   0.07   \nl
S0a$-$Sbc &   0.71   &   0.08  &   0.76   \nl
Sc$\leq$  &   0.40   &   0.16  &   0.29   \nl
\enddata
\tablenotetext{a}{The probability that rejects the null hypothesis.}
\end{deluxetable}

%----------------------------------------------------------------------------
\clearpage
%%%%%%%%%%%%%%%%%%%%% TABLE 6 %%%%%%%%%%%%%%%%%%%%%%%%%%%%%
\begin{deluxetable}{ccccccccc}
  \tablenum{6}
  \tablewidth{0pt}
  \tablecaption{\ Frequency distributions of the nuclear activity types 
                between the HCGs and the field\tablenotemark{a}
               }
  \tablehead{
    %%% HEAD 1 %%%
    &
    \multicolumn{2}{c}{All\tablenotemark{b}} &
    \multicolumn{2}{c}{E-S0}&
    \multicolumn{2}{c}{S0a-Sbc} &
    \multicolumn{2}{c}{Sc$\leq$} \nl
    %%% HEAD 2 %%%
    &
    \colhead{HCGs}  &
    \colhead{Field} &
    \colhead{HCGs}  &
    \colhead{Field} &
    \colhead{HCGs}  &
    \colhead{Field} &
    \colhead{HCGs}  &
    \colhead{Field}
  }

\startdata
AGN        & 44.4 (28) & 43.7 (167) & 28.0  (7) & 60.8  (59) &
             70.8 (17) & 57.2  (83) & 28.6  (4) & 17.9  (25)  \nl

\ion{H}{2} & 25.4 (16) & 45.5 (174) &  8.0  (2) &  4.1   (4) &
             16.7  (4) & 40.0  (58) & 71.4 (10) & 80.0 (112) \nl

Abs        & 30.2 (19) & 10.7  (41) & 64.0 (16) & 35.1  (34) &
             12.5  (3) &  2.8   (4) &  0.0 (0)  &  2.1   (3) \nl
\hline
P($\chi^{2}$)\tablenotemark{c} & 
\multicolumn{2}{c}{$3.93\times10^{-5}$} &
\multicolumn{2}{c}{0.01} & 
\multicolumn{2}{c}{0.01} &
\multicolumn{2}{c}{0.55}

\tablenotetext{a}{Numbers in parentheses are the actual numbers.}
\tablenotetext{b}{``All'' means that the total sample of E$-$S0, 
                  S0a$-$Sbc, and Sc$\leq$ galaxies.}
\tablenotetext{c}{The possibility that rejects the null hypothesis.}
\enddata
\end{deluxetable}

%----------------------------------------------------------------------------
\clearpage
%%%%%%%%%%%%%%%%%%%%% TABLE 7 %%%%%%%%%%%%%%%%%%%%%%%%%%%%%
\begin{deluxetable}{ccccccccc}
  \tablenum{7}
  \tablewidth{0pt}
  \tablecaption{\ Frequency distributions of the morphological types 
                between the HCGs and the field as a function of 
                the nuclear activity\tablenotemark{a}
               }
  \tablehead{
    %%% HEAD 1 %%%
    &
    \multicolumn{2}{c}{All\tablenotemark{b}} &
    \multicolumn{2}{c}{AGN} &
    \multicolumn{2}{c}{\ion{H}{2}} &
    \multicolumn{2}{c}{Abs} \nl
    %%% HEAD 2 %%%
    &
    \colhead{HCGs}  &
    \colhead{Field} &
    \colhead{HCGs}  &
    \colhead{Field} &
    \colhead{HCGs}  &
    \colhead{Field} &
    \colhead{HCGs}  &
    \colhead{Field}
  }

\startdata
E-S0     & 39.7 (25) & 25.4  (97) & 25.0  (7) & 35.3 (59) & 
           12.5  (2) &  2.3   (4) & 84.2 (16) & 82.9 (34) \nl
S0a-Sbc  & 38.1 (24) & 38.0 (145) & 60.7 (17) & 49.7 (83) &
           25.0  (4) & 33.3  (58) & 15.8  (3) &  9.8  (4) \nl
Sc$\leq$ & 22.2 (14) & 36.6 (140) & 14.3  (4) & 15.0 (25) &
           62.5 (10) & 64.4 (112) &  0.0  (0) &  7.3  (3) \nl
\hline
P($\chi^{2}$)\tablenotemark{c} & 
\multicolumn{2}{c}{0.03} &
\multicolumn{2}{c}{0.51} & 
\multicolumn{2}{c}{0.08} &
\multicolumn{2}{c}{0.41}

\tablenotetext{a}{Numbers in parentheses are the actual numbers.}
\tablenotetext{b}{``All'' means that the total sample of AGN, \ion{H}{2},
                  and absorption galaxies.}
\tablenotetext{c}{The possibility that rejects the null hypothesis.}

\enddata
\end{deluxetable}

%----------------------------------------------------------------------------
\clearpage

%%%%%%%%%%%%%%%%%%%%% TABLE 8 %%%%%%%%%%%%%%%%%%%%%%%%%%%%%
\begin{deluxetable}{cccc}
  \tablenum{8}
  \tablewidth{35em}
  \tablecaption{\ Comparisons of expected numbers of the activity 
                types with the observations for the HCG 
                sample\tablenotemark{a}
               }
  \tablehead{
    %%% HEAD 1 %%%
    &
    \colhead{Observations} &
    \colhead{Corrected numbers\tablenotemark{b}} &
    \colhead{Expected numbers} 
  }

\startdata
AGN        & 44.4 (28)  & 56.5 (35) & 49.2 (31) \nl
\ion{H}{2} & 25.4 (16)  & 25.4 (16) & 34.9 (22) \nl
Abs        & 30.2 (19)  & 19.4 (12) & 15.9 (10) \nl
\hline
P($\chi^{2}$)\tablenotemark{c} & 0.14 & 0.50 &  

\tablenotetext{a}{Numbers in parentheses are the actual numbers.}
\tablenotetext{b}{Expected numbers under assuming that we have taken 
AGN for Abs for a half (3/7) of early-type galaxies}
\tablenotetext{c}{The possibility that rejects the null hypothesis.}
\enddata
\end{deluxetable}

%----------------------------------------------------------------------------